\documentclass[11pt]{article} 
\usepackage{hyperref}
\usepackage[T1]{fontenc}
\usepackage{bm}
\usepackage{graphicx}

\newcommand{\be}{ \begin{equation}}
\newcommand{\ee}{\end{equation}} 
\usepackage{amsmath}
\usepackage{amsfonts}
\usepackage[T1]{fontenc}
\usepackage{bm}
\usepackage{appendix}
\usepackage{amssymb}

\begin{document} 
\def\theequation{\arabic{section}.\arabic{equation}} 
\begin{titlepage} 
\title{Unveiling the physics of partial differential equations with 
heuristics} 
\author{Valerio Faraoni\\ 
{\small \it  Department of Physics \& Astronomy, Bishop's University}\\ 
{\small \it 2600 College Street, Sherbrooke, Qu\'ebec Canada 
J1M~1Z7} 
}
\date{} \maketitle 
\thispagestyle{empty} 
\vspace*{1truecm} 
\begin{abstract} 

Heuristic arguments and order of magnitude estimates for partial 
differential equations highlight essential features of the physics they 
describe. We present order of magnitude estimates, and their limitations, 
for the three classic second order PDEs of mathematical physics (wave, 
heat, and Laplace equations), for first order transport equations, and for 
two non-linear wave equations. It is beneficial to expose the beginning 
student to these considerations before jumping into more rigorous 
mathematics. Yet these simple arguments are missing from physics 
textbooks.

\end{abstract} 
\vspace*{1truecm} 
\begin{center} 
Preprint of an article submitted for consideration in 
\href{https://www.worldscientific.com/worldscinet/tpe?adgroupid=23075846775&campaignid=336010695&creative=81823290255&gclid=EAIaIQobChMIgeOa3eXG-AIVoMLCBB0myAgFEAAYASAAEgKP8PD_BwE&hsa_acc=8603214540&hsa_ad=81823290255&hsa_cam=336010695&hsa_grp=23075846775&hsa_kw=&hsa_mt=&hsa_net=adwords&hsa_src=g&hsa_tgt=dsa-146675683935&hsa_ver=3&keyword=&utm_campaign=Dynamic+Search+Ads&utm_medium=ppc&utm_source=adwords&utm_term=}{{\em 
The Physics Educator}} (World Scientific Publishing Company)
\end{center} 
\end{titlepage}

\def\theequation{\arabic{section}.\arabic{equation}}


\section{Introduction} 
\label{sec:1}
\setcounter{equation}{0}

Typically, undergraduate physics students are introduced to the classic 
second order linear partial differential equations (PDEs) of mathematical 
physics through various courses in the curriculum, maybe taking a course 
devoted to this subject, but only in upper years. Beginning students 
usually 
struggle with PDEs, learning to make sense of them, and how to use them, 
only after substantial effort. After some practice with the three main 
classes of second order PDEs of mathematical physics, {\em i.e.}, 
hyperbolic, parabolic, and elliptic equations exemplified by the wave, 
diffusion, and Laplace equations, respectively, some ``sloppy math'' 
considerations often arise, which are simple order of magnitude estimates 
facilitating the understanding of this subject. The early discussion of 
these observations when first introducing PDEs would be more beneficial. 
These order of magnitude estimates capture essential features of these 
PDEs, which are later rediscovered in specific analytical solutions ({\em 
e.g.}, Refs.\cite{Courant,WeberArfken}).

The classic wave, heat, and Laplace PDEs embody the essential features of 
the physical phenomena they describe, which are pillars of an education in 
physics, and it is best to show the student 
this physical content before the details.  Most mathematics textbooks do 
not present the derivation of these PDEs from basic physical principles 
(conservation of energy for the heat equation, Newton's second law for 
points of a string displaced from equilibrium for the wave equation, {\em 
etc.}). Even when such derivations are presented, many students fail to 
realize that the corresponding PDEs exhibit features that have a certain 
degree of universality. These PDEs describe phenomena that go beyond, say, 
the wave in a string, and they are important as prototypes for all linear 
transport, diffusion, or wave-like phenomena. Simple hand-waving 
introductions to PDEs do not find a place in mathematics or even 
physics texbooks but are 
nevertheless very useful for the physics student and should be included in 
physics courses in some way. Here we summarize semi-qualitative aspects of 
PDEs that, while certainly apparent to instructors, are probably reputed 
to be beneath dignified textbooks. Nevertheless, they are greatly 
appreciated by beginning students in introductory lectures.

\section{Linear PDEs} 
\label{sec:2}
\setcounter{equation}{0}

Let us begin with the one-way transport equations in one spatial dimension
\begin{equation}
\frac{\partial u}{\partial t} \mp c \, \frac{\partial u}{\partial x}=0 \,, 
\label{mptransport}
\end{equation}
where $c$ is a constant coefficient and, for definiteness, we consider 
infinite domains $-\infty < x < +\infty$ in this section. Let $T$ and $L$ 
be, respectively, the time scale and spatial scale of 
variation of the quantity $u(t,x)$ that is being transported. Clearly, the 
coefficient $c$ describes the speed at which the quantity $u$ is being 
transported. The order of magnitude estimates $u_t \sim u/T$ and $u_x \sim 
u/L$  (where $u_t \equiv \partial u/\partial t$, $ u_x\equiv \partial 
u/\partial x$, {\em etc.}) give
\begin{equation}
\frac{u}{T} \mp c\, \frac{u}{L} \sim 0 
\end{equation}
from which it follows that 
\begin{equation}
L= \pm c \, T \,.
\end{equation}
Thus, already at first sight, Eq.~(\ref{mptransport}) says that some 
quantity $u$ is being transported at speed $c$, covering the 
distance $L$ in the time $T$, to the right if the upper 
sign is  chosen or to the left if the lower sign is applied. This 
physical intuition motivates guessing the solutions
\begin{equation}
u_1(t,x) = f(x+ct) \,, \quad \quad u_2(t,x) = g (x-ct) 
\end{equation}
where $f$ and $g$ are regular functions of their arguments, say continuous 
with their first derivatives. It is then straightforward to verify that 
$u_1(t,x)$ describes transport to the left (since $c \sim -L/T$) and 
$u_2(t,x)$ transport to the right (since $c\sim + L/T$). These solutions 
are just the translations $f(\xi)$, $g(\eta)$ of static profiles $f(x)$ 
and $g(x)$ to the left or to the right, respectively, with $x\to \xi 
\equiv x+ct$ or $x\to \eta \equiv x-ct$. These Galilean translations  
express the motion of the whole functions along the line at constant speed 
to the left or to the right, and hold thanks to the constancy of the 
coefficient $c$ (there are also physically meaningful situations in which 
$c$ 
can be considered approximately constant over  a certain region, or in 
which $c$ varies slowly, which leads to meaningful approximations).

Let us proceed with the one-dimensional homogeneous heat or diffusion 
equation
\begin{equation}
\frac{\partial u}{\partial t} =a \nabla^2 u  \label{heateq}
\end{equation}
describing heat conduction in the absence of sources, where $a$ is the 
Fourier coefficient (here assumed to be constant)  and 
$ u(t,x)$ is the temperature. Denote with $T$ the time scale of 
variation of $u$ and with $L$  its spatial scale of variation. We 
have, in order of magnitude,  
\begin{equation}
\frac{\partial u}{\partial t} \sim \frac{u}{T} \,, \quad\quad 
\frac{\partial u}{\partial x^i} \sim \frac{u}{L } \,, \quad\quad 
\nabla^2 u \sim \frac{\partial^2 u}{\partial (x^i)^2} \sim \frac{u}{L^2} 
\,;
\end{equation}
then the heat equation~(\ref{heateq}) gives  
$$
\frac{u}{T} \sim a \, \frac{u}{L^2} 
$$
and 
\begin{equation}
L=\sqrt{a \, T}
\end{equation}
This simple equation provides valuable physical insight: it expresses the 
characteristic feature of a random walk 
that 
the macroscopic distance travelled grows with the square root of time, if 
one thinks of $u(t, \vec{x}) $ as something that  spreads in a random 
way. 
In a random walk, the root mean square distance travelled by the average  
particle grows with the square root of the number $N$ of scatterings it 
experiences, which is a signature of the stochastic nature of diffusion.  
$N$ is proportional to the time $t$ expired from the beginning of the 
walk, so 
$x_\mathrm{rms} \propto \sqrt{N}\sim \sqrt{D \, t}$, where $D$ is the 
diffusion coefficient analogous to the Fourier coefficient for heat 
conduction. The simple order of magnitude estimate in the PDE reveals 
an essential feature of the physical process it 
describes. (The fact that $x_\mathrm{rms}=\sqrt{Dt}$ is 
rediscovered by the student in the Gaussian solution of the heat 
equation,\cite{WeberArfken} which is the basis for the many Gaussian 
plume models 
used to describe the spreading of pollutants in environmental 
physics.\cite{BoekerGrondelle}).

Continuing with qualitative considerations, in the heat 
equation~(\ref{heateq}) the speed of diffusion $\partial u/\partial t $ of 
the heat is proportional to the spatial ``curvature'' $\nabla^2 u$ of the 
temperature, which is more intuitive in one dimension where $\nabla^2u$ 
reduces to $d^2 u/d x^2$. In first year calculus, students are 
instructed to regard $d^2y/dx^2 $ as a measure of the curvature of the 
graph of the function $y(x)$ and they can easily relate to it.  (See 
Ref.~\cite{Styer} for a more nuanced discussion of the meaning of the 
Laplacian.) The larger this curvature, the faster the process. Finally, 
the fundamentally irreversible nature of diffusion is highlighted by the 
non-invariance of Eq.~(\ref{heateq}) under the time reversal $t\to -t$.

Pass now to the wave equation
\begin{equation}
\nabla^2 u - \frac{1}{c^2} \, \frac{\partial^2 u}{\partial t^2}=0 
\label{waveeq}
\end{equation}
with constant coefficient $c$, for which the same order of magnitude 
reasoning employing time scale 
$T$ and length scale $L$ of variation of the solution $u(t, \vec{x})$ 
gives
$$
\frac{u}{L^2} \sim  \frac{1}{c^2} \, \frac{u}{T^2} 
$$
and
\begin{equation}
L=c \, T  \,,
\end{equation}
which is nothing but the relation 
\begin{equation}
c=\lambda \, \nu \label{v=lambdanu}
\end{equation}
between wavelength $\lambda$, frequency $\nu$, and speed $c$ of a wave, to 
which a student relates from algebra-based elementary physics courses 
(which, however, do not discuss Eq.~(\ref{waveeq})). The naive order of 
magnitude 
approach tells the student that something travels with speed $c$ and 
Eq.~(\ref{v=lambdanu}) reports the basic fact that distance 
covered~$=$~velocity~$\times$~time. The wave $u(t,x)$ travels the distance 
$L$ in the time $T = \nu^{-1}$ (the period), and this is the way the 
relation $c=\lambda \nu$ is introduced in elementary physics courses.

At this point in the introduction of the wave PDE, it is fruitful to 
focus on the one-dimensional wave 
equation
\begin{equation}
\frac{\partial^2 u}{\partial x^2}   - \frac{1}{c^2} \, \frac{\partial^2 
u}{\partial t^2}=0 \,,
\end{equation}
and derive the general solution on the infinite line 
\begin{equation}
u(t,x) = f(x+ct)+g(x-ct) \,,
\end{equation}
where $f$ and $g$ are regular functions of their arguments (say, 
continuous with their first and second order derivatives). By 
formally splitting the d'Alembert operator as
\begin{equation}
\Box \equiv \frac{\partial^2 }{\partial x^2}   - \frac{1}{c^2} \, 
\frac{\partial^2 }{\partial t^2}= 
\left( \frac{\partial }{\partial x} + \frac{1}{c} \, 
\frac{\partial }{\partial t} \right) 
\left( \frac{\partial }{\partial x}  - \frac{1}{c} \, 
\frac{\partial }{\partial t} \right) \,,
\end{equation}
the second order one-dimensional wave equation for $u(t,x)$ is equivalent 
to the set of two first order PDEs
\begin{eqnarray}
\frac{\partial u}{\partial x}   + \frac{1}{c} \, 
\frac{\partial u}{\partial t} &=& 0 \,,\\
&&\nonumber\\
\frac{\partial u}{\partial x}   - \frac{1}{c} \, 
\frac{\partial u}{\partial t} &=& 0 \,.
\end{eqnarray}
These are transport equations of the form~(\ref{mptransport}) already 
discussed and help making sense of the fact that the general solution of 
the wave equation on the infinite line describes waves travelling to the 
left or to the right. In this discussion, the pulses $f(x+ct)$ and 
$g(x-ct)$ maintain their profiles unchanged as they 
propagate, hence they describe waves that do not suffer 
dispersion.\cite{Main,Pain,Klemens} 

Let us come now to the prototypical elliptic equation, the Laplace 
equation
\begin{equation}
\nabla^2 u=0 
\end{equation}
that describes static phenomena. At first sight, the order of magnitude 
reasoning based on the scale of variation $L$ of $u(\vec{x})$ would 
produce 
$u/L^2\simeq 0$, or $L\sim \infty$, which is not particularly 
enlightening and could lead one to conclude that, since nothing travels in 
this case, the 
comparison with the heat and the wave equations is fruitless. However, 
one can do better and note that, contrary to the case of the previous two 
equations, one can see the Laplace equation as a (first order) equation 
{\em for the 
gradient of $u$}, instead of $u$ itself. It is often convenient to 
simplify the situation and look at the special 
case of one spatial dimension, in which a PDE is as simple as possible. 
The Laplace equation becomes trivial in one dimension,
\begin{equation}
\frac{d^2u}{dx^2}=0 \,,
\end{equation}
and the solution is a straight line. The meaning is that the solutions of 
the Laplace equation are ``as straight as possible'', which reproduces 
also the property that the solutions assume maxima and minima on the 
boundary of the domain of integration. These properties survive in higher 
dimension (this approach is used, {\em e.g.}, in 
Ref.\cite{Griffiths} with great benefit for the student).

In one spatial dimension, the gradient reduces to $u'(x) \equiv du/dx$ and 
the 
Laplace equation {\em for $u'$} reads 
\begin{equation} 
\frac{du'}{dx}=0 \,, 
\end{equation} 
expressing the fact that the slope $u'$ of the solution $u(x)$ is constant 
or, the graph of the solution is straight and has no curvature $u''$. With 
this view, the naive order of magnitude approach gives $u'/L \simeq 0$ or, 
the length scale of variation of $u'$ is infinite. Then $u'$ does not 
change: the slope is constant, which means that the graph of the solution 
$u$ has no curvature.

Moving to two dimensions, $u(x,y)$ can vary in two independent 
directions $x$ and $y$, but these variations are not independent: the 
Laplace equation links them. If the 
curvature of $u$ is positive in the $x$-direction, $u_{xx}>0$, 
it must simultaneously be negative in the $y$-direction, $u_{yy}<0$, to 
compensate so that the total curvature $ u_{xx}+u_{yy}$ of $u$ 
vanishes identically. This behaviour is exemplified by the harmonic 
function 
\begin{equation}
v(x,y)=x^2-y^2 \,,
\end{equation}
for which $v_{xx}=2 $ but $v_{yy}=-2$ so that the sum $v_{xx}+v_{yy}$ 
vanishes. Another revealing harmonic function is 
\begin{equation}
w(x,y) = xy
\end{equation}
which is linear in each of the independent $x$- and $y$-  
directions: $w_x=y$ and $w_y=x$, then $w_{xx}=w_{yy}=0$ satisfying 
the Laplace equation.

Another elliptic equation that the physics student is bound to encounter  
is the Helmoltz equation 
\begin{equation}
\nabla^2 u +k^2 u =0 \,.
\end{equation} 
Let us refer, specifically, to the problem of solving the wave 
equation by separation of variables, in which the Helmoltz equation 
appears.\cite{Courant,WeberArfken} Again, it is useful to first comment 
on its one-dimensional version 
\begin{equation}
\frac{d^2u}{dx^2}+k^2 u(x)=0 \,,
\end{equation}
which is nothing but the harmonic oscillator equation describing 
oscillations in the variable $x$ ({\em i.e.}, in space) with angular 
frequency $k$. By 
extension, when solving the wave equation by separation of variables, the 
higher-dimensional Helmoltz equation describes oscillations of the 
spatial part $u$ of the wave in space 
(in the direction of the wave vector $\vec{k}$).  
The order of magnitude estimate using the 
spatial scale $L$ of variation of the solution $u$ 
gives 
\begin{equation} 
k \sim \frac{1}{L} \,, 
\end{equation} 
which is the order of magnitude version of the relation $k=2\pi / \lambda$ 
between wave vector $k$ and wavelength $\lambda$ for the propagating 
waves.

\section{Beyond linear}
\label{sec:3}
\setcounter{equation}{0}

The naive order of magnitude estimate still serves well the student who 
encounters much more complicated non-linear equations. To this regard, 
standard physics courses only discuss linear, small-amplitude waves 
without mentioning the existence of non-linear waves. They ignore what a 
trip to the beach reveals, {\em i.e.}, large-amplitude waves in shallow 
water crashing on the shore. Many undergraduate physics students end their 
course of studies believing that all waves are linear. It is more honest 
to 
mention the existence of non-linear waves in introductory courses, and 
then restrict the discussion to small-amplitude waves instead of hiding 
this beautiful subject forever for fear of the mathematics involved. 
Later on, non-linear waves can still be made accessible to 
undergraduates.\cite{Malfliet}

As an example of the order of magnitude technique applied to non-linear 
PDEs, consider the Burgers equation
\begin{equation}
\frac{\partial u}{\partial t} = u \, \frac{\partial u}{\partial x}  
\label{Burgerseq}
\end{equation}
describing solitonic waves. Based on what the student now knows about  
transport equations, it is tempting to view Eq.~(\ref{Burgerseq}) as a 
transport equation of the form\begin{equation}
\frac{\partial u}{\partial t} = c(u)  \, \frac{\partial u}{\partial x} 
\end{equation}
where the velocity $c(u)=u$ depends on 
the wave's amplitude according to $c = u_t/u_x =u$. The order of 
magnitude 
estimate using temporal and spatial scales of variation $T$ and $L$ 
of the solution $u$ still gives 
\begin{equation}
c= \frac{u_t}{u_x} \sim \frac{ u/T}{u/L} = \frac{L}{T} \,,
\end{equation}
expressing the fact that something is still being transported, but now 
this ratio is a function of the amplitude $u$ of the wave and is smaller 
for low-amplitude waves and larger for large-amplitude ones. 
Large-amplitude waves travel faster ({\em i.e.}, the dispersion depends on 
the wave's amplitude), and this dependence makes the Burgers equation 
non-linear.

Consider now the Korteweg-de Vries (KdV) equation describing 
one-dimensional non-linear waves in 
a shallow water channel. Let $\tau $ and $\rho $ be the surface tension 
and density of water, respectively, while $h$ is the depth of the channel, 
$g$ is the acceleration of gravity, and $u(t,x)$ is the vertical 
displacement of the water surface from its position of equilibrium. The 
third order KdV equation is 
\begin{equation}
\frac{\partial u}{\partial t} + c \, \frac{\partial u}{\partial x} 
+\gamma\, u \, \frac{\partial u}{\partial x} +\epsilon \, \frac{\partial 
^3 u}{\partial x^3}=0 \,, \label{KdV}
\end{equation}
where
\begin{equation}
\gamma = \frac{3c}{2h} 
\end{equation}
is a parameter describing non-linearity, 
\begin{equation}
\epsilon =  c \left( \frac{h^2}{6}- \frac{\tau}{2\rho \, g} \right)
\end{equation}
is a parameter describing dispersion, and $c$ is the velocity of linear 
(low-amplitude) waves, {\em i.e.}, the velocity that waves would have if 
the terms weighted by the parameters $\gamma$ and $\epsilon$ were absent. 
In fact, if $\gamma=\epsilon=0$, the KdV equation~(\ref{KdV}) reduces to 
the transport equation~(\ref{mptransport}) already discussed, describing 
transport to the left.

Applying again the order of magnitude reasoning with scales of variation 
of the solution $u(t,x)$ in time and space $T$ and $L$, we have now
\begin{equation}
\frac{u}{T} \sim  - \left[ \left( c +\gamma \, u \right) \frac{u}{L} 
+\epsilon \, \frac{u}{L^3} \right] 
\end{equation}
(the negative sign describing transport to the left) and the velocity of 
propagation is approximately
\begin{equation}
\frac{L}{T} \sim - \left( c+ \gamma \, u + \frac{\epsilon}{L^2} \right) 
= -c \left[ 1+\frac{3u}{2h} +\frac{h^2}{6L^2}-\frac{\tau}{2\rho \, g L^2} 
\right] \equiv 
-c \left[ 1+\frac{3u}{2h} +\frac{h^2}{6L^2}-\frac{\ell_c^2}{L^2} 
\right] \,, \label{KdVsimplified}
\end{equation}
where $\ell_c \equiv \sqrt{\frac{\tau}{\rho \, g}}$ is the capillary 
length used in the discussion of dispersion.\cite{Main,Pain,Klemens}  
Equation~(\ref{KdVsimplified}) makes it clear that the speed of the waves 
depends on their amplitude as well as on the ratio between wave amplitude 
and depth of the channel (longer waves ``feel'' the depth of the channel, 
which is shallow for them) and the ratio between capillary length and 
``wavelength''. Because of the unintuitive physics, now it is 
not 
possible to justify analytical solutions as done for the linear wave 
equation, but they are still accessible to the 
undergraduate.\cite{Malfliet,HansenNicholson}

\section{Limitations of the approach}
\label{sec:4}
\setcounter{equation}{0}

By definition, heuristic explanations, order of magnitude estimates, and 
qualitative arguments are not rigorous. Their usefulness stems from the 
fact that they provide intuition and physical insight before going through 
a rigorous course and lengthy chains of arguments. Beginning students 
benefit from glimpsing key physical features before digesting an entire 
semester of theorems, proofs, corollaries, and exercises, not only in 
terms of motivation but also because these discoveries facilitate the 
understanding of rigorous results and make them more likely to 
be remembered. While extremely valuable, these glimpses should be tempered 
by an analysis of their limitations, which we address in this section.

The first issue is that the order of magnitude estimates exposed in the 
previous section use the notion of characteristic scale, which is 
extremely valuable and widely used in physics and engineering. In general, 
however, it is not simple to define the concept of characteristic scale 
for physical quantities, nor is there a unique definition. In fact, if a 
physical quantity is expressed by a Fourier representation with many 
wavenumbers, a characteristic scale is not even defined unambiguously. 
This is probably the reason why this concept is avoided in introductory 
courses, yet one is throwing out the baby with the bathwater by doing so.

A convenient (but not unique) approach to defining a characteristic length 
for quantities satifying linear differential equations is based on 
their Fourier decomposition. It consists of 
introducing, say, the characteristic length scale $L $ of a physical 
quantity $ u(t, x) $ at time $t$ using the average of its wavenumber 
power spectrum ${\cal P}(t, k)$,
\begin{equation}
L \equiv  \frac{2\pi}{ \langle k \rangle} 
\end{equation}
where 
\begin{equation}
\langle k \rangle = \int_0^{+\infty} dk\, {\cal P}(t, k) k \,.
\end{equation}
Here 
\begin{equation} 
{\cal P}(t, k) = a |u(t, k)|^2 \,,
\end{equation}
$a>0$ is a normalization constant, and 
\begin{equation} 
u(t, k) = \frac{1}{\sqrt{2\pi}} \int_{-\infty}^{+\infty} 
dx\, u(t,x) \, \mbox{e}^{ikx}
\end{equation}
is the Fourier coefficient at time $t$ of the Fourier component 
with wavenumber $k$. Similarly, a characteristic time scale at the 
position 
$x$  can be introduced as the average of the frequency power spectrum 
${\cal P}(x, \omega)$,
\begin{equation}
T \equiv \frac{2\pi}{ \langle \omega \rangle} \,,
\end{equation}
where ${\cal P}(x, \omega)= b |u(x, \omega)|^2$.  These definitions are 
intuitive if the spectrum is peaked around a 
particular wavelength (or wavenumber) and less meaningful otherwise.  It 
is clear that, with these definitions, the concept of length scale 
depends on the time and that of time scale on the position, which 
complicates matters. Nevertheless, even though these scales are only 
defined {\em locally}, the concept is still useful because differential 
equations describe physics locally. There is merit, therefore, in 
referring to ``characteristic scales'' when introducing PDEs, invoking 
intuition rather than entering elaborate discussions about the Fourier 
representation of physical quantities. This discussion can be postponed to 
a time when students have already become familiar with Fourier 
representations and linear PDEs.

A second issue is that characteristic scales are defined in relation with 
initial (for evolution PDEs) and boundary conditions, which may have their 
own characteristic scales. Initial conditions may have length or time 
scales associated with them, and the same can be said of boundary 
conditions on a finite domain (the size of a finite domain is itself a 
scale and shows up in the solutions of the PDE, for example waves in a 
string clamped at both ends or the wavefunction for the quantum particle 
in a box). 
Initial-boundary value problems for PDEs and their well-posedness 
are considered in the presence of initial and boundary conditions, which 
is certainly mentioned in introductions to PDEs.  Initial and boundary 
conditions, as well as possible variable source terms, may affect 
characteristic scales.

One can summarize as follows the conditions under which the heuristic 
arguments of the previous sections are valid:

\begin{itemize}

\item PDEs are considered on an infinite domain (in one dimension, 
$-\infty <x < +\infty$). This assumption guarantees that no scale is 
associated with the size of a finite domain (or, in higher dimension, its 
shape).

\item The PDEs considered have constant coefficients. This assumption may 
be replaced by assuming that these coefficients are approximately constant 
over a region of interest, as done many times in the physics curriculum. 
This assumption avoids introducing  characteritic scales through these  
coefficients. For example, in the 
one-dimensional Schr\"odinger equation for a particle of mass $m$ 
described by the wave function $\psi(t,x)$
\begin{equation}
i\hbar \, \frac{\partial \psi}{\partial t}= -\frac{\hbar^2}{2m} \, 
\frac{\partial^2\psi}{\partial^2 x} + V(x) \, \psi \,,
\end{equation}
the potential energy $V(x)$, in general, introduces a spatial scale.

\item The PDE is homogeneous or else the source terms  do not vary in time 
and space. When this is true, no scales are introduced by the sources.

\item If the PDE describes time evolution, the initial conditions 
contain only 
one length or time scale, {\em i.e.}, its Fourier spectrum is peaked on a 
rather narrow range of frequencies or wavenumbers. In this case, if a 
physical quantity $u(t, x)$ is Fourier-decomposed as
\begin{equation}
u(t,x) = \frac{1}{\sqrt{2\pi}} \int_{-\infty}^{+\infty} dk \, u(t, k) \,,
\end{equation}
then 
\begin{equation}
\frac{\partial u}{\partial x} = 
\frac{1}{\sqrt{2\pi}} \int_{-\infty}^{+\infty} dk \, ik \, u(k) \,,
\end{equation}
and 
\begin{equation}
\frac{u}{L} \approx ik_0 u 
\end{equation} 
provided that the spectrum peaks around the wavenumber $k_0$. 
\end{itemize}

Finally, if the PDE considered is linear, it is satisfied by single 
Fourier modes of the physical quantity $u(t,x)$ obeying it, which has 
well-defined scales associated with them. Non-linearities cause mode-mode 
mixing and the presence of multiple scales when two incommensurate wave 
numbers $k_1$ and $k_2$ mix. This is the case, for example, of forward or 
inverse cascades in fluid dynamics, where energy is transferred from large 
to small scales, or {\em vice-versa}, respectively.

\section{Discussion}
\label{sec:5}
\setcounter{equation}{0}

Using order of magnitude estimates for PDEs, as suggested in the previous 
sections, is common practice in dimensional analysis\cite{Huntley67} in 
engineering and applied physics (see, {\em e.g.}, 
Refs.\cite{BoekerGrondelle,CampbellNorman}), or in fluid mechanics and 
oceanography\cite{Defant61,PickardEmery,PondPickard}, where many terms are 
dropped from the relevant long PDEs, retaining only those that dominate in 
certain physical regimes. Therefore, there is no real reason for 
withdrawing this kind of analysis from the introductory discussion of PDEs 
when beginner students could instead benefit from them. Appealing to 
mathematical rigour does not justify hiding physical intuition which is 
very useful as an entry point into PDEs and even more for understanding 
rigorous results derived later in the course. Indeed, using heuristic 
arguments and order of magnitude estimates for PDEs with constant 
coefficients early on has the extra advantage that intuitive features of 
the physical processes described are rediscovered later in specific 
analytical solutions of these PDEs, reinforcing the point. As usual, 
heuristic arguments have limitations, which we have highlighted in 
Sec.~\ref{sec:4}, but their benefits outweight these limitations.

\section*{Acknowledgments}

This work is supported by Bishop's University.


\end{document}